\documentclass[conference]{IEEEtran}
\usepackage{amsmath}
\usepackage{amssymb}
\usepackage{floatfig,epsfig}
\usepackage{subfigure}
\usepackage{verbatim}
\usepackage{pifont}
\usepackage{algorithm}
\usepackage{algorithmic}
\usepackage{alltt}
\usepackage{bm}
\usepackage{isolatin1}
\usepackage{graphics}
\usepackage{epic}
\usepackage{eepic}
\usepackage{graphpap}
\usepackage{cite}
\usepackage{psfrag}
\usepackage{rotating}

\newtheorem{pro}{Proposition}

\newtheorem{lem}{Lemma}

\newcommand{\vect}[1]{{\mathbf{#1}}}

\renewcommand{\a}{\vect{a}}

\ifCLASSINFOpdf
\else
\fi

\begin{document}

\title{Exact Optimized-cost Repair  in Multi-hop Distributed Storage Networks}

\author{Majid Gerami, Ming Xiao
\\  Communication Theory Lab, Royal Institute of Technology, KTH, Sweden, \\ E-mail: \{gerami, mingx\}@kth.se\\}
\date{\today}


\maketitle

\begin{abstract}
\normalsize
The problem of exact repair of a failed node in multi-hop networked distributed storage systems is considered. Contrary to the most of the current studies which model the repair process by the direct links from surviving nodes to the new node, the repair is modeled by considering the multi-hop network structure, and taking into account that there might not exist direct links from all the surviving nodes to the new node. In the repair problem of these systems, surviving nodes may cooperate to transmit the repair traffic to the new node. In this setting, we define the total number of packets transmitted between nodes as repair-cost. A lower bound of the repair-cost can thus be found by cut-set bound analysis. In this paper, we show that the lower bound of the repair-cost is achievable for the exact repair of MDS codes in tandem and grid networks, thus resulting in the minimum-cost exact MDS codes. Further, two suboptimal (achievable) bounds for the large scale grid networks are proposed.
\end{abstract}

\begin{keywords}
\noindent Network coding, Distributed storage systems, Exact repair.
\end{keywords}
%

\IEEEpeerreviewmaketitle

\section{Introduction}\label{sec:intro}
As node failure  in distributed storage systems is norm, the process of regenerating a new node, known as repair process, has extensively been studied (see, e.g., [1]-[9] and references therein). In the repair process an enough number of surviving nodes transmit packets to the new node. The repair process in turn follows with the cost in bandwidth and energy of transmission or in another aspect the cost of reading data from storage nodes. To reduce the cost from these aspects, \cite{Dimk01}  proposes regenerating codes, having optimal repair bandwidth;  \cite{Pap01} proposes locally repairable codes, thereby the new node connects less number of storage nodes; and \cite{El01} proposes fractional repetition codes, thereby the new node reads less from each storage node. To reduce encoding/decoding complexity and communication overhead, exact repair has been studied in \cite{Rashmi}, \cite{Suh01}. In the exact repair, the content of the new node is exactly the same as the failed node, while in the functional repair  the content of the new node might be different, but the new node beside surviving nodes preserve the regenerating code property \cite{Dimk01}.

In the most of the current studies the graph in the repair is depicted by the links directly connecting  surviving nodes to the new node. This might not be a proper model in a distributed storage system built on a multi-hop network. In other words, there  might not exist direct links from some of the surviving nodes to the new node. In this setting, several surviving nodes may cooperate to transmit the repair traffic to the new node. Assuming one packets transmission from a surviving node to its neighbor node costs  one unit of cost (e.g. one unit of energy), we define the total number of  transmissions between nodes as repair-cost. This definition based on the considering the network structure is expected to be a better representation of the transmission cost in the repair process of multi-hop networks.

In this paper, we investigate the exact MDS regenerating codes having optimized repair-cost in multi-hop distributed storage systems. To this aim, the remainder of the paper is organized as follows. In Section \ref{sec:MinCost}, we briefly describe the method of finding the lower bound of the repair-cost. Then in Section \ref{sec:CodeConstruction}, we propose the exact MDS regenerating codes achieving the lower bound of the repair-cost in tandem networks and in a $2\times 3$ grid network. Further, in Section \ref{sec:Supoptimal} we propose two achievable bounds of the repair-cost for the exact MDS codes in large scale grid networks.

\section{ A Lower bound of the repair cost in  multi-hop networks}\label{sec:MinCost}

 Consider a distributed storage system with $n$ storage nodes where nodes are connected together with an arbitrary topology. A file of size $M$ bits, is equally divided to $k$ parts and then coded by an $(n,k)$-MDS code such that every node stores $M/k$ bits and every $k$ nodes can rebuild the original file. When a node fails, a new node is regenerated by the help of $d$ number of surviving nodes. The information flow graph for a repair can be represented by an acyclic directed graph $\mathcal{G}=(\mathcal{N,A})$, where $\mathcal{N}$ denotes the set of nodes with cardinality $|\mathcal{N}|=n$ and $\mathcal{A}$ is the set of links in the network. There is a cost $c_{ij}$  associated for the use of  each link $(ij)\in \mathcal{A}$. For simplicity, in this paper we assume transmission on each link on the network costs one unit. That is $c_{ij}=1$ for $(ij)\in \mathcal{A}$. Thus, the total repair cost, denoted by $\sigma_c$, depends only on the total repair traffic on the links. Let $z_{ij}$ denotes number of packets transmitted from node $i$ to node $j$ in the repair process. Hence, the repair-cost is formulated as,
 \begin{equation}
 \sigma_c=\Sigma_{(ij)\in \mathcal{A}} z_{ij}.
 \end{equation}

A lower bound of $\sigma_c$ is derived by minimizing the repair-cost over $z_{ij}$s such that all the cuts of connecting DC to the new node and any set of $k-1$ out of $n-1$ surviving nodes have to be greater than $M$, the file size. The objective function and constraints in this optimization problem are linear. Consequently the lower bound is derived by solving a linear programming problem. The solution also provides a vector, namely subgraph, containing the values of $z_{ij}$s for the lower bound of the repair cost. We note that in this paper we derive the lower bounds for $d= n-1$, except otherwise stated.

For the notation, we use bold lower case letters for vectors and bold upper case letters for matrices in the following sections.


\section{Explicit Construction of Exact Optimal-cost Codes}\label{sec:CodeConstruction}

In this section, we show that the lower bound of the repair cost can be achieved for the exact repair in the examples of tandem and grid networks. The code design has two steps. First,  the lower bound of the repair cost and the corresponding  subgraph are found by the method briefly described in Section \ref{sec:MinCost} (for more details please see \cite{Maj01J}). In the second step, the exact MDS regenerating code matching to the corresponding subgraph is derived.

\subsection{Explicit Code Construction in  Tandem Networks}\label{sec:CodeConstructionTandem}
In a tandem network, nodes are in a line topology.  That is, each
node is linked to two neighboring nodes, except the nodes in the line ends, which have only one neighbor. When a node fails and a new
node joins, the repair traffic is relayed by intermediate nodes
to the new node. For these networks, the lower bound of the repair-cost is derived by the following proposition.

\begin{pro} Consider a tandem distributed storage network consisting of
$n$ storage nodes which storing a file with size $M$ packets (fragments). Each node stores $M/k$ fragments and every $k$
nodes can reconstruct the original file ($\mathrm{MDS}$ property). The repair-cost is greater than or equal to $M$, i.e.,

\begin{equation}
\sigma_c \geqslant M.
\label{tandem_eq}
\end{equation}
Moreover, the minimum repair cost is by cooperation of the $k$ nearest
surviving nodes to the new node where each node transmits $M/k$ fragments to its neighbor.
\label{pro:TandemCost}\end{pro}
\proof [Proof]{Proof is omitted due to space limitation. Please find the details in \cite{Maj01J}.}

 In what follows, we shall give the code construction with the optimal cost. We split the source file of a size $M$ into $k$ fragments. We denote the source file by vector $\mathbf{m}=[m_1 m_2 \cdots m_k]^T$. To construct $(n,k)\texttt{-MDS}$ code, we use a $k \times n$ Vandermonde matrix  $G$ as a generator matrix.
\begin{equation}
\mathbf{G}=  \left(\begin{array}{cccc}
1 & 1 & \cdots & 1\\
\alpha_1 & \alpha_2 &   \cdots & \alpha_n \\
\alpha_1^2 & \alpha_2^2 & \cdots & \alpha_n^2 \\
\vdots & \vdots & \ddots & \vdots\\
\alpha_1^{k-1} & \alpha_2^{k-1} &  \cdots & \alpha_n^{k-1}
\end{array}\right),\end{equation}
where  $\alpha_i$s for $i \in \{ 1,\cdots, n\}$ are distinct elements from a finite field $\mathrm{GF(q)}$.

By the property of Vandermonde matrix, every $k \times k$ submatrix of $\mathbf{G}$ is full rank if $\alpha_i$s, $(i \in \{ 1,\cdots, n\})$, are distinct elements. This requires $q > n$. Each column e.g., column $i$ $(i \in \{1,\cdots,n\})$ in matrix $\mathbf{G}$ represents the code in node $i$. We denote the coded data in node $i$ as $\mathbf{v_i}$, then,
\vspace{-6 mm}

 \begin{equation}
 v_{i}=m_1+m_2 \alpha_{i}+\cdots+m_k \alpha_{i}^{k-1}= [1\text{ } \alpha_t \text{ } \cdots \alpha_i^{k-1}] \mathbf{m}.
\end{equation}
Clearly, a data collector can reconstruct the source file by connecting to any $k$ nodes. We shall further show that the  minimum-cost exact repair  is possible by linear codes. For illustration, we assume nodes are labeled in order, i.e., node $1$ connects to node $2$, node $2$ connects to node $1$ and $3$, and so on. By Proposition  \ref{pro:TandemCost}, for $M=k$ fragments, the lower bound of repair-cost is by transmitting $M/k=k/k=1$ fragment to the neighbor. Assume node $t$ $(t \in \{1,\cdots,n\})$ fails and a set of nodes $\{ \texttt{node}_{t-k_{1}},\texttt{node}_{t-k_{1}+1},\cdots, \texttt{node}_{t-1} \}$ and a set of nodes $\{ \texttt{node}_{t+1},\texttt{node}_{t+2},\cdots, \texttt{node}_{t+k_{2}} \}$ $(k_{1}+k_{2}=k)$ help to regenerate the new node.

The repair process is as follows. The new node receives fragments from two directions, namely, aggregated data from nodes $\texttt{node}_{t-k_{1}}$ via \texttt{node} $t-1$,  and  aggregated data from nodes $\texttt{node}_{t+k_{2}}$ via \texttt{node} $t+1$. Thus, in one direction node $\texttt{node}_{t-k_{1}}$ multiplies its content by a coefficient $\xi_{t-k_{1}}$ from $GF(q)$ and sends the result to node $\texttt{node}_{t-k_{1}+1}$. Then node $\texttt{node}_{t-k_{1}+1}$, multiplies its content by $\xi_{t-k_{1}+1}$ and  combines the result to the received fragment and then sends its combined fragment to its next neighbor $\texttt{node}_{t-k_{1}+2}$.  Finally  \texttt{node} $t-1$  transmits the combined fragment $w_{t-1}$, which is
\begin{equation}\begin{array}{ll}
w_{t-1}=\xi_{t-k_{1}} v_{t-k_{1}}+\xi_{t-k_{1}+1} v_{t-k_{1}+1}+ \cdots + \xi_{t-1} v_{t-1}\\
=\xi_{t-k_{1}}(m_1+m_2 \alpha_{t-k_{1}}+\cdots+m_k \alpha_{t-k_{1}}^{k-1})\\
+\xi_{t-k_{1}+1}(m_1+m_2 \alpha_{t-k_{1}+1}+\cdots+m_k \alpha_{t-k_{1}+1}^{k-1})\\
+\cdots+\xi_{t-1}(m_1+m_2 \alpha_{t-1}+\cdots+m_k \alpha_{t-1}^{k-1}).
\end{array}\end{equation}
In another direction, node \texttt{node} $t+1$ similarly sends the aggregated fragment to the new node. That is  \texttt{node} $t+1$  transmits the combined fragment $w_{t+1}$, which is,
\begin{equation}\begin{array}{ll}
w_{t+1}=\xi_{t+k_{2}} v_{t+k_{2}}+\xi_{t+k_{2}-1} v_{t+k_{2}-2}+ \cdots + \xi_{t+1} v_{t+1}\\
=\xi_{t+k_{2}}(m_1+m_2 \alpha_{t+k_{2}}+\cdots+m_k \alpha_{t+k_{2}}^{k-1})\\
+\xi_{t+k_{2}-1}(m_1+m_2 \alpha_{t+k_{2}-1}+\cdots+m_k \alpha_{t+k_{2}-1}^{k-1})\\
+\cdots+\xi_{t+1}(m_1+m_2 \alpha_{t+1}+\cdots+m_k \alpha_{t+1}^{k-1}).
\end{array}\end{equation}
To achieve exact repair, we set $w_{t-1}+w_{t+1}=v_t$
\begin{equation}
 v_{t}=m_1+m_2 \alpha_{t}+\cdots+m_k \alpha_{t}^{k-1}.
\end{equation}
Thus, vector $\underline{\xi}=[\xi_{t-k_{1}},  \cdots, \xi_{t-1}, \xi_{t+1},  \cdots, \xi_{t+k_{2}}]^{T}$ should be selected such that,
\begin{equation}
  \underbrace{\begin{pmatrix}
1 & 1 & \cdots & 1\\
\alpha_{t-k_1} & \alpha_{t-k_1+1} &   \cdots & \alpha_{t+k_2} \\
\alpha_{t-k_1}^2 & \alpha_{t-k_1+1}^2 & \cdots & \alpha_{t+k_2}^2 \\
\vdots & \vdots & \ddots & \vdots\\
\alpha_{t-k_1}^{k-1} & \alpha_{t-k_1+1}^{k-1} &  \cdots & \alpha_{t+k_2}^{k-1}
\end{pmatrix}}_{\mathbf{A}} \begin{pmatrix} \xi_{t-k_{1}} \\ \xi_{t-k_{1}+1}\\ \vdots \\ \xi_{t+k_{2}}\end{pmatrix}=
 \begin{pmatrix}
 1\\
 \alpha_{t}\\
  \alpha_{t}^2\\
   \vdots\\
     \alpha_{t}^{k-1}\end{pmatrix}.\end{equation}
Since  matrix $\mathbf{A}$ is non-singular, we can  select linear codes for repair at node $t$ as  $\mathbf{\xi_t}=\mathbf{A}^{-1}\mathbf{v_t}$ that makes the exact repair possible.  Hence, for the successful reconstruction and repair process, a finite field  $q > n$ suffices.

\vspace{0.2cm}
\vspace{-6 mm}
\subsection{Explicit  Construction for the  minimum-cost exact repair in a $2 \times 3$ Grid Network}\label{sec:CodeConstructionGrid}
 In what follows, we show that the lower bound of the repair-cost is also achievable for the exact repair in a $2 \times 3$ grid network. We use systematic codes. That is, a file is equally divided to $k$ parts ($k$ uncoded fragments). Then, $k$ systematic nodes store the uncoded fragments. Other parity nodes store a linear combination of the original fragments such that  for any selection of $k$ nodes,  the original file can be reconstructed (MDS property). We design a code such that  all systematic nodes can be exactly regenerated with the minimum cost.   We note that in our codes, parity nodes can also be repaired with the minimum-cost, following the dual relationship between parity and systematic nodes, as proposed in \cite{Suh01}. Thus, we only show the exact minimum-cost repair on systematic nodes in what follows.

\begin{figure}
 \centering
 \psfrag{a}[][][3]{ $\alpha$ }
 \psfrag{node1}[][][3]{ node 1 }
 \psfrag{node2}[][][3]{ node 2 }
 \psfrag{node3}[][][3]{ node 3 }
 \psfrag{node4}[][][3]{ node 4 }
 \psfrag{node5}[][][3]{ node 5 }
 \psfrag{node6}[][][3]{ node 6 }
 \psfrag{n11}[][][3]{ $\rho_1 a_2+\xi_{11} a_1+\xi_{12} b_1+\xi_{13} c_1$ }
 \psfrag{n12}[][][3]{ $\xi_{11} a_2+\xi_{12}+\xi_{13}c_2$ }
 \psfrag{n21}[][][3]{ $\rho_2 b_2+\xi_{21} a_1+\xi_{22} b_1+\xi_{23} c_1$ }
 \psfrag{n22}[][][3]{ $\xi_{21} a_2+\xi_{22} b_2+\xi_{23} c_2$ }
 \psfrag{n31}[][][3]{ $\rho_3 c_2+\xi_{31} a_1+\xi_{32} b_1+\xi_{33} c_1$ }
 \psfrag{n32}[][][3]{ $\xi_{31} a_2+\xi_{32} b_2+\xi_{33} c_2$ }
 \psfrag{n41}[][][3]{ $a_1$ }
 \psfrag{n42}[][][3]{ $a_2$ }
 \psfrag{n51}[][][3]{ $b_1$ }
 \psfrag{n52}[][][3]{ $b_2$ }
 \psfrag{n61}[][][3]{ $c_1$ }
 \psfrag{n62}[][][3]{ $c_2$ }
  \psfrag{f451}[][][3]{  $a_1$ }
 \psfrag{f25}[][][3]{ $\rho_2 b_2+\xi_{21} a_1+\xi_{22} b_1+\xi_{23} c_1$  }
 \psfrag{f36}[][][3]{ $ \rho_3 c_2+\xi_{31} a_1+\xi_{32} b_1+\xi_{33} c_1$ }
 \psfrag{f562}[][][3]{ $c_1$ }
 \psfrag{f561}[][][3]{ $\xi_{31} a_1+\xi_{32} b_1$ }
 \resizebox{9cm}{!}{\epsfbox{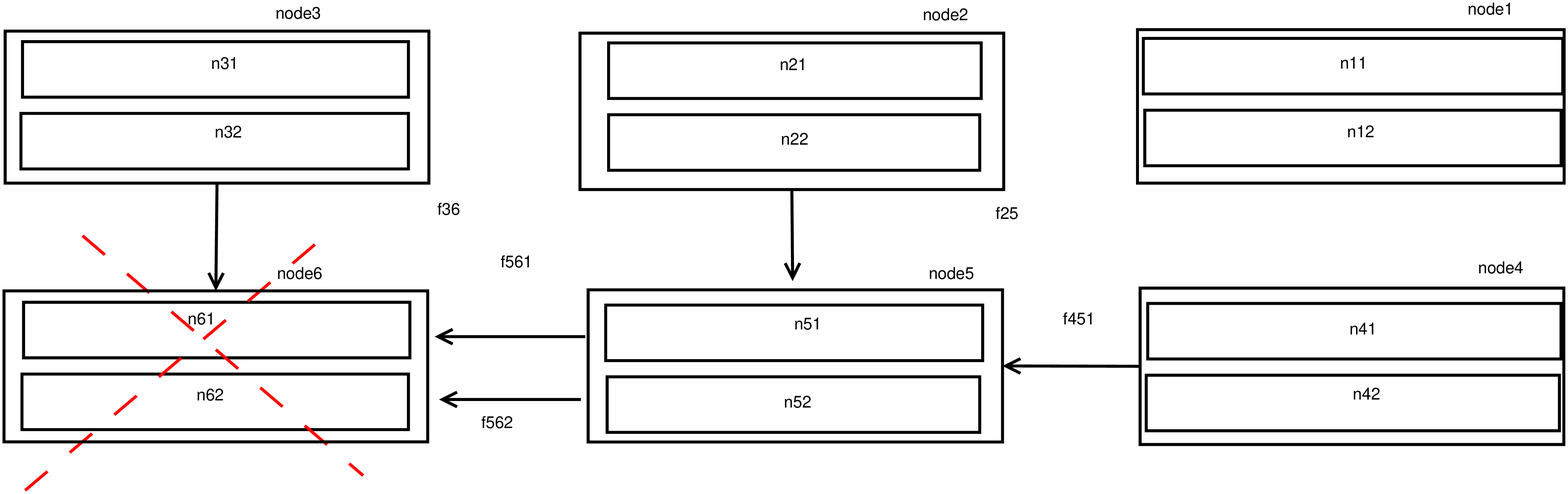}}
\caption{Exact and optimal-cost repair in the $2\times 3$  grid network using interference alignment/cancellation technique.}
 \label{ExactRepairGrid}
\end{figure}

Consider a distributed storage system with parameters $(n=6, k=3, M=6)$, where each node stores $M/k=2$ fragments, in  a $2\times 3$ grid network shown in Fig.  \ref{ExactRepairGrid}. When a node fails a new node is regenerated by the remaining $5$ nodes. The code has $3$ nodes as systematic nodes and $3$ nodes as parity nodes. By next proposition, the minimum repair-cost is $5$ units. The corresponding optimal-cost subgraph  for different nodes might be different. The optimal-cost repair for the repair on node  $6$ has been shown in the Fig. \ref{ExactRepairGrid}. 

\begin{pro} The lower bound of  repair cost for systematic nodes $4$, $5$, and $6$ in the distributed storage system with parameters $(M=6, k=3)$, where each node stores $M/k=2$ fragments, in a $2\times 3$ grid network having one unit of cost for transmitting one fragment between neighboring nodes, is by $5$ units. The corresponding minimum-cost subgraphs $\mathbf{z}=(z_{(12)}, z_{(14)}, z_{(23)}, z_{(25)}, z_{(36)},
z_{(54)}, z_{(65)})$ for the repair of nodes $4$, $5$, and $6$ are respectively
$(0, 1, 0, 1, 0, 2, 1)$ , $(0, 0, 0, 1, 1, 1, 2)$, and $(0, 1, 0, 1, 0, 2, 1)$.\end{pro}
\proof[Proof (sketch)]{ For the repair on each node, we perform $\binom{5}{2}=10$ cut-set analysis on the information flow graph, as in \cite{Maj01J}, and then solve a linear programming problem to achieve the minimum-cost subgraphs}.

For this $2\times 3$ grid network, we show the existence of  linear codes for the exact repair in the systematic nodes with minimum-cost. The exact regenerating code is found by the interference alignment/cancellation techniques, as will be described later.
\subsubsection{Code Construction}
For the source file of size $M=6$ fragments, we put the fragments into two vectors as $\mathbf{m_1}=[a_1, b_1, c_1]$ and $\mathbf{m_2}=[a_2, b_2, c_2]$. Every node then stores two fragments.   Nodes $4$, $5$, and $6$ are systematic nodes. That is, node $4$ stores fragments $a_1$, and $a_2$, node $5$ stores fragments $b_1$, and $b_2$, and node $6$ stores fragments $c_1$, and $c_2$.

For parity nodes, let $p_{i1}$ and $p_{i2}$ denote two fragments stored on node $i$ for $i=1,2,3$. To construct a proper code, we first construct a $3\times 3$ Vandermonde matrix as
\[ \mathbf{v}= \begin{pmatrix} \mathbf{\xi_1} &\mathbf{\xi_2} & \mathbf{\xi_3} \end{pmatrix}=\begin{pmatrix}\xi_{11} &\xi_{21} & \xi_{31}  \\ \xi_{12} &\xi_{22} & \xi_{32}  \\\xi_{13} &\xi_{23} & \xi_{33}  \\  \end{pmatrix}\] \[=\begin{pmatrix} 1 &1 &1  \\ \alpha_1 & \alpha_2  &\alpha_3  \\\alpha_1^2 & \alpha_2^2  &\alpha_3^2   \end{pmatrix}, \]
where $\alpha_1, \alpha_2, \alpha_3  $ are distinct elements in $\mathrm{GF(q)}$. Parameter $q$ denotes the finite field size and is the design parameter. For the Vandermonde matrix there must be $q \geq 3$. Finally, the fragments on nodes $1$, $2$ and $3$ are coded as,
\begin{equation} \begin{pmatrix} p_{11}\\p_{12}\end{pmatrix}= \begin{pmatrix} \rho_1 a_2+\mathbf{\xi_1}\mathbf{m_1}\\ \mathbf{\xi_1}\mathbf{m_2} \end{pmatrix}, \end{equation}
 \begin{equation} \begin{pmatrix} p_{21}\\p_{22}\end{pmatrix}= \begin{pmatrix} \rho_2 b_2+\mathbf{\xi_2}\mathbf{m_1}\\ \mathbf{\xi_2}\mathbf{m_2} \end{pmatrix}, \end{equation}
 \begin{equation} \begin{pmatrix} p_{31}\\p_{32}\end{pmatrix}= \begin{pmatrix} \rho_3 c_2+\mathbf{\xi_3}\mathbf{m_1}\\ \mathbf{\xi_3}\mathbf{m_2}, \end{pmatrix}, \end{equation}
where  $\rho_1,\rho_2,\rho_3 $ are non-zero elements in $\mathrm{GF(q)}$.

\subsubsection{Exact repair}
We show the repair process on node $6$ and $5$. Other nodes are similar.
 \begin{itemize}
 \item {Exact repair for node $6$:}  First, \texttt{node $4$} transmits  fragment $a_1$ to the \texttt{node $5$}. Then \texttt{node $2$} sends $p_{21}$ to the \texttt{node $5$}. Then \texttt{node $5$} removes the \emph{interference} $a_1, b_1, b_2$ from $p_{21}$ and sends $c_1$ to the new node. Then \texttt{node $4$} sends $\xi_{31}a_1+\xi_{32}b_2$ which is aligned with the interference in $p_{31}$. Finally, the new node receives $c_1$ and  $c_2+\xi_{31}a_1+\xi_{32}b_1+\xi_{33}c_1$, and an interference vector $\xi_{31}a_1+\xi_{32}b_1$.  After removing  interference, $c_1$ and $c_2$ are recovered in the new node.
 \item {Exact repair on node $5$:} First, \texttt{node $3$}  transmits  fragment $p_{31}$ to \texttt{node $6$}. In $\texttt{node 6}$,  interference $c_1$ and $c_2$ are removed and the fragment $\xi_{31}a_1+\xi_{32}b_1$ is transmitted to \texttt{node $5$}. \texttt{Node $5$} also receives fragment $a_1$ from node 4. Then interference $a_1$ is also removed from $\xi_{31}a_1+\xi_{32}b_1$ and $b_1$ is recovered. To recover $b_2$, \texttt{node $2$} transmits $p_{21}$. Then \texttt{node $6$} transmits $c_1$. We note that the new node already received fragment $a_1$. Thus interference $a_1$ and $c_1$ are removed from fragment $p_{21}$. By recovering $b_1$, the fragment $b_2$ is recovered from fragment $p_{21}.$

 \end{itemize}
 Thus, with $\rho_1,\rho_2,\rho_3\neq 0$, and Vandermonde matrix $\mathbf{\xi}$, we   achieve exact repair with minimum-cost.
\subsubsection{Reconstruction process}
   We show as follows that  MDS property can be preserved if coefficients $\rho_1,\rho_2,\rho_3$ are non-zero, and matrix $\mathbf{\xi}$ is a Vandermonde matrix. We consider the MDS property in four different cases: (1) $\mathrm{DC}$ connects to $3$ parity nodes, (2) $\mathrm{DC}$ connects to $2$ parity node and $1$ systematic nodes, (3) $\mathrm{DC}$ connects to $1$ parity node and $2$ systematic nodes, and (4) $\mathrm{DC}$ connects to $3$ systematic nodes.
\begin{itemize}
\item Case 1: There will be only one way of selecting $3$ parity nodes.  For the $\mathrm{DC}$ to recover the file, it requires to solve the following linear equations.
     \begin{equation}\underbrace{\begin{pmatrix}      \xi_{11} &\xi_{12} &\xi_{12} &\rho_1 &0 & 0 \\ \xi_{21} &\xi_{22} &\xi_{23} &0 &\rho_2 & 0\\\xi_{31} &\xi_{32} &\xi_{33} &0 &0 & \rho_3 \\0 &0 &0  &\xi_{11} &\xi_{12} &\xi_{12}\\0 &0 &0  &\xi_{21} &\xi_{22} &\xi_{23}\\0 &0 &0  &\xi_{31} &\xi_{32} &\xi_{32}\\
     \end{pmatrix}}_{\mathbf{F}} \begin{pmatrix} a_1\\ b_1\\c_1\\a_2\\b2\\c_2\end{pmatrix}=\mathbf{r},\end{equation}
where vector $\mathbf{r}$ is the received vector and is known for the $\mathrm{DC}$. For recovering the original file, it requires matrix $\mathbf{F}$ to be invertible. This can be proven by the following lemma.
\lem[Binomial inverse theorem] {For the square matrices $\mathbf{A}$, $\mathbf{B}$ and $\mathbf{C}$, when $\mathbf{A}=\mathbf{B}+\mathbf{C}$, then \begin{equation} \mathbf{A}^{-1}=(\mathbf{B}+\mathbf{C})^{-1}=\mathbf{B}^{-1}-\mathbf{B}^{-1}(\mathbf{I}+\mathbf{C}\mathbf{B}^{-1})^{-1}\mathbf{C}\mathbf{B}^{-1},\end{equation} where $\mathbf{I}$ is a proper identity matrix.} For the case,
  \[\mathbf{F}=\underbrace{\begin{pmatrix}      \xi_{11} &\xi_{12} &\xi_{12} &0 &0 & 0 \\ \xi_{21} &\xi_{22} &\xi_{23} &0 &0 & 0\\\xi_{31} &\xi_{32} &\xi_{33} &0 &0 & 0 \\0 &0 &0  &\xi_{11} &\xi_{12} &\xi_{12}\\0 &0 &0  &\xi_{21} &\xi_{22} &\xi_{23}\\0 &0 &0  &\xi_{31} &\xi_{32} &\xi_{32}\\
     \end{pmatrix}}_{\mathbf{B}}+\]\begin{equation}\underbrace{\begin{pmatrix}     0 &0 &0 &\rho_1 &0 & 0 \\ 0 &0 &0 &0 &\rho_2 & 0\\0 &0 &0 &0 &0 & \rho_3 \\0 &0 &0  &0 &0 &0\\0 &0 &0  &0 &0 &0\\0 &0 &0  &0 &0 &0\\
          \end{pmatrix}}_{\mathbf{C}}\end{equation}
     By Lemma 1, since matrices $\mathbf{B}$ and $\mathbf{I}+\mathbf{C}\mathbf{B}^{-1}$ are invertible, matrix $\mathbf{F}$ is invertible. Thus, the DC can rebuild the source.
     \item Case 2: $\mathrm{DC}$ connects to $2$ parity nodes and $1$ systematic node. There will be $9$ scenarios of selecting $2$ parity nodes among $3$ parity nodes and $1$ systematic node among $3$ systematic nodes. We use one of them for illustration. Assume $\mathrm{DC}$ connects to nodes $1$, $2$, and $4$. Thus $\mathrm{DC}$ has access to $a_1$, $a_2$. To recover the original file $\mathrm{DC}$ solves the following linear equation.
          \begin{equation} \begin{pmatrix}  \xi_{12} &\xi_{13} &0 & 0 \\ \xi_{22} &\xi_{23} &\rho_2 & 0\\0 &0 &\xi_{12} &\xi_{13} \\0 &0 &\xi_{22} &\xi_{23}
     \end{pmatrix} \begin{pmatrix} b_1\\ c_1\\b_2\\c_2\end{pmatrix}=\mathbf{r^{'}}, \end{equation}
     where vector $\mathbf{r}'$ is known for $\mathrm{DC}$. Using Lemma 1, the linear equation has unique solution and then $\mathrm{DC}$ recovers the original file.
     \item Case 3: $\mathrm{DC}$ connects to $1$ parity node and $2$ systematic nodes. For this scenario,  $\mathrm{DC}$ recovers the original file with given  Vandermonde matrix $\xi$ and $\rho_1,\rho_2,\rho_3\neq 0$.
      \item Case 4: $\mathrm{DC}$ connects to $3$ systematic nodes. In this scenario, it is straightforward for $\mathrm{DC}$ to recover the source.
\end{itemize}
Hence, above codes have minimum-cost exact repair  with a finite field size $q\geq 3$.

We have previously shown the minimum-cost exact regenerating codes in tandem networks (for arbitrary parameters $n, k$) and in a grid network for a special case of $2 \times 3$ grid network. We still cannot prove that the lower bound of the repair-cost  is achievable for the exact repair in a general case (a network with arbitrary topology and large number of nodes). In addition, deriving the lower bound  has exponential complexity and requires $\binom{n-1}{k-1}$ cut-set bound analysis.  This motivates us to  investigate some achievable bounds (suboptimal-cost methods) for MDS exact regenerating codes in large scale networks. This will be described in next section.

\section{Suboptimal-cost Exact regenerating codes in multi-hop networks}\label{sec:Supoptimal}
 In this section, we propose two suboptimal approaches based on the optimal-cost results found in the tandem and $2\times 3$ grid networks. From the optimal-cost approach in tandem networks, we find that there are always suboptimal approaches for $(n,k)$ MDS codes by neglecting some available links, and considering a set of nodes as tandem structure. That is, by finding  $k$ nearest nodes to the new node. Consequently, the new node can be regenerated  with the cost of $M$, where any node participating in the repair process transmits $M/k$ fragments to its neighbor. We call the approach \texttt{Suboptimal$\#1$} method. The second suboptimal approach is motivated from the result of optimal-cost repair in the $2 \times 3$ grid network. We generalize the results to networks with $(n=2k, k, d=k+1)$ grid topology as follows. We call the approach as \texttt{Suboptimal$\#2$}  method.
\subsection{Suboptimal-cost repair in an $r \times s $ grid network}
For general ($n=2k,k,d=k+1$) grid networks, the code construction is similar to the $2\times 3$  network. A file of size $M$ is divided to $2k$ fragments. Then two vectors $\mathbf{m_1}$ and $\mathbf{m_2}$ denote the original file, where $\mathbf{m_1}=\begin{pmatrix} m_{11}, m_{12}, \cdots, m_{1k} \end{pmatrix}$, and $\mathbf{m_2}=\begin{pmatrix} m_{21}, m_{22}, \cdots, m_{2k} \end{pmatrix}$. There are $k$ systematic nodes and $k$ parity nodes.  For the $ r \times s$ grid network, total number of nodes are $n=rs=2k$ nodes, hence number of rows ($r$) or column ($s$) are even integer (at-least one of them). Without loss of generality, assume $s$ is an even integer. Let \texttt{$node_{ij}$} denotes the node on row $i$ and column $j$.  We put the systematic nodes on columns $1$ to $s/2$, and rows $1$ to $r$.   We store two fragments in each node. Let $a_{ij}^1$ and $a_{ij}^2$ denote the systematic data on the $node_{ij}$ for $i=1,\cdots,r$ and $j=1,\cdots,s/2$. Then $a_{ij}^1=m_{1t}$, and $a_{ij}^2=m_{2t}$, where $t=(i-1)s/2+j$. In words, each systematic node stores one fragment from $\mathbf{m_1}$ and one fragment from $\mathbf{m_2}$. Corresponding to each systematic node we assign a parity node  which locates on the same row, but on column $j+s/2$. For the parity nodes, we use Vandermonde matrices for encoding. Thus, we first construct a $k\times k$ Vandermonde matrix and denote as $\mathbf{\xi}$. Then, each column of matrix $\mathbf{\xi}$, denoted as $\xi_i$ for $i=1,\cdots,k$, is used for each parity node. We use column vector $\xi_t$, $t=(i-1)s/2+j$, for encoding the  parity node on row $i$ and column $j$. If $p_{ij}^1$ and $p_{ij}^2$ denote the parity fragments stored on node $ij$,  then $p_{ij}^1=a_{ij}^2+\mathbf{\xi_t} \mathbf{m_1}$ and $p_{ij}^2=\mathbf{\xi_t} \mathbf{m_2}$. We further show that the exact repair and and reconstruction processes can successfully be done for the proposed code.

\begin{figure}[h!]
 \centering
 \psfrag{a}[][][1.5]{ $\alpha$ }
 \psfrag{node1}[][][2]{ node 1 }
 \psfrag{node2}[][][2]{ node 2 }
 \psfrag{node3}[][][2]{ node 3 }
 \psfrag{node4}[][][2]{ node 4 }
 \psfrag{node5}[][][2]{ node 5 }
 \psfrag{node6}[][][2]{ node 6 }
 \psfrag{p111}[][][2]{ $a_2+\mathbf{\xi_1} \mathbf{m_1}$ }
 \psfrag{p112}[][][2]{ $\mathbf{\xi_1} \mathbf{m_2}$ }
 \psfrag{p211}[][][2]{ $b_2+\mathbf{\xi_2} \mathbf{m_1}$ }
 \psfrag{p212}[][][2]{ $\mathbf{\xi_2} \mathbf{m_2}$ }
 \psfrag{p311}[][][2]{ $c_2+\mathbf{\xi_3} \mathbf{m_1}$ }
 \psfrag{p312}[][][2]{ $\mathbf{\xi_3} \mathbf{m_2}$ }
 \psfrag{p411}[][][2]{ $d_2+\mathbf{\xi_4} \mathbf{m_1}$ }
 \psfrag{p412}[][][2]{ $\mathbf{\xi_4} \mathbf{m_2}$ }
 \psfrag{p121}[][][2]{ $e_2+\mathbf{\xi_5} \mathbf{m_1}$ }
 \psfrag{p122}[][][2]{ $\mathbf{\xi_5} \mathbf{m_2}$ }
 \psfrag{p221}[][][2]{ $f_2+\mathbf{\xi_6} \mathbf{m_1}$ }
 \psfrag{p222}[][][2]{ $\mathbf{\xi_6} \mathbf{m_2}$ }
 \psfrag{p321}[][][2]{ $g_2+\mathbf{\xi_7} \mathbf{m_1}$ }
 \psfrag{p322}[][][2]{ $\mathbf{\xi_7} \mathbf{m_2}$ }
 \psfrag{p421}[][][2]{ $h2_2+\mathbf{\xi_8} \mathbf{m_1}$ }
 \psfrag{p422}[][][2]{ $\mathbf{\xi_8} \mathbf{m_2}$ }
 \psfrag{a111}[][][2]{ $a_1$ }
 \psfrag{a112}[][][2]{ $a_2$ }
 \psfrag{a211}[][][2]{ $b_1$ }
 \psfrag{a212}[][][2]{ $b_2$ }
 \psfrag{a311}[][][2]{ $c_1$ }
 \psfrag{a312}[][][2]{ $c_2$ }
 \psfrag{a411}[][][2]{ $d_1$ }
 \psfrag{a412}[][][2]{ $d_2$ }
 \psfrag{a121}[][][2]{ $e_1$ }
 \psfrag{a122}[][][2]{ $e_2$ }
 \psfrag{a221}[][][2]{ $f_1$ }
 \psfrag{a222}[][][2]{ $f_2$ }
 \psfrag{a321}[][][2]{ $g_1$ }
 \psfrag{a322}[][][2]{ $g_2$ }
 \psfrag{a421}[][][2]{ $h_1$ }
 \psfrag{a422}[][][2]{ $h_2$ }
 \psfrag{fp11}[][][2]{ $a_2+\mathbf{\xi_1} \mathbf{m_1}$ }
 \psfrag{fp12}[][][2]{ $\xi_{11} a_1+\xi_{12} b_1+\xi_{13} c_1+\xi_{14} d_1+\xi_{16} f_1+\xi_{17} g_1+\xi_{18} h_1$ }
 \psfrag{fp22}[][][2]{ $\xi_{21} a_1+\xi_{22} b_1+\xi_{23} c_1+\xi_{24} d_1+\xi_{25} e_1+\xi_{27} g_1+\xi_{28} h_1$ }
 \psfrag{f211}[][][2]{ $\xi_{21} a_1+\xi_{15} e_1$ }
 \psfrag{f212}[][][2]{ $\xi_{12} b_1+\xi_{13} c_1+\xi_{14} d_1+\xi_{16} f_1+\xi_{17} g_1+\xi_{18} h_1$ }
 \psfrag{f431}[][][2]{ $h_1$ }
 \psfrag{f321}[][][2]{ $\xi_{13} c_1+\xi_{14} d_1+\xi_{16} f_1+\xi_{17} g_1+\xi_{18} h_1$ }
 \psfrag{f322}[][][2]{ $\xi_{23} c_1+\xi_{24} d_1+\xi_{25} e_1+\xi_{27} g_1+\xi_{28} h_1$ }
\resizebox{9cm}{!}{\epsfbox{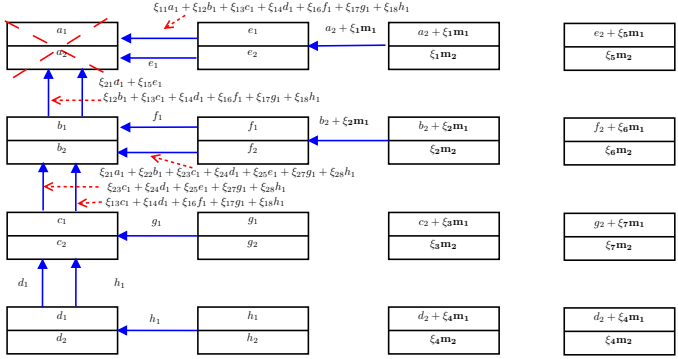}}
\caption{Exact and suboptimal-cost repair in the $4\times 4$  grid network using interference alignment/cancellation technique.}
 \label{ExactRepairGrid_2}
\end{figure}

\subsubsection{Exact regenerating process}  For the exact regeneration of the systematic nodes on row $i$ and column $j$, all other systematic nodes, and parity nodes $p_{ij}$ and $p_{(i+1)j}$ shall cooperate.  Assume a systematic node with the fragments $a_{ij}^1$ and $a_{ij}^2$ fails. All other systematic nodes transmit one of their fragments to the new node. The parity node $node_{i(j+s/2)}$ transmits the fragment $a_{ij}^2+\mathbf{\xi_{t}} \mathbf{m1}$ to the new node. A parity node $node_{(i+1)(j+s/2)}$ transmits $a_{(i+1)j}^2+\mathbf{\xi_{t}}\mathbf{m_1}$ to node $node_{(i+1)j}$. The interference are removed from the parity fragments through the the path  toward the new node or at the new node. Consequently, there would be two independent equations with two unknown fragments, $a_{ij}^1$ and $a_{ij}^2$ in the new node. Thus the fragment on $node_{ij}$ is recovered. As an example the code and the exact repair processes on $node_{11}$ in a $4\times 4$ grid network have been shown in Fig.  \ref{ExactRepairGrid_2}. Thus, the exact repair is achieved.
\subsubsection{Reconstruction process}  The parity nodes are encoded following a Vandermonde matrix. Similar to the approach for the $2\times 3$ grid network and using  Lemma 1, we can verify that the reconstruction  is feasible for selecting any $k$ out of $n$ nodes. Thus, for reconstruction $q > k$ suffices.
\vspace{-5 mm}
\subsection{Numerical Results}
  In what follows, we shall show the performance gain of our approaches with increasing nodes in the network. We apply the suboptimal approaches in large networks. We compare those suboptimal approaches with the approach which models the repair process by direct links and then optimizing the repair-cost (denoted as \texttt{Method\#3}). We compare the repair cost for several grid network as shown in Table \ref{Tabel:CompareMethods}. In this table the gain of suboptimal approaches in reducing the repair cost has been shown. We can see that a  larger number of nodes in the network results in a larger gain for reducing the repair-cost.
  \begin{table}
  \caption{Comparing the repair-cost by different methods}
  \begin{center}
     \begin{tabular}{c c c c }
     \hline \hline
     Network dimension & Suboptimal\#1 & Suboptimal\#2 & Method\#3 \\ \hline
     $2\times 2$ & 4 & 4 & 4 \\ \hline
     $2\times 3$ & 6 & 6 & 5 \\ \hline
     $2\times 4$ & 8 & 8 & 7 \\ \hline
     $3\times 4$ & 12 & 10 & 13 \\ \hline
     $4\times 4$ & 16 & 14 & 21 \\ \hline
     $5\times 4$ & 20 & 16 & 29 \\ [1ex] \hline
     \end{tabular}
 \end{center}
 \label{Tabel:CompareMethods}
 \end{table}

\section{Conclusions} \label{sec:conclusion}
We proposed an explicit construction of the optimal-cost exact  repair in the general case of tandem  networks. By interference alignment/cancellation techniques we proposed an explicit code for the exact repair in a $2\times 3$ grid network. For the grid network with large number of nodes,  two achievable repair-cost for the exact MDS regenerating codes have been evaluated. Our analysis shows that considering network topology in the exact repair problem is beneficial in reducing the repair-cost.

\end{document}